\documentclass[12pt]{article}
\usepackage{amsmath,amssymb}%,showkeys}
\newtheorem{theorem}{Theorem}[section]
\newtheorem{proposition}[theorem]{Proposition}
\newtheorem{lemma}{Lemma}
\newtheorem{corollary}[theorem]{Corollary}
\newtheorem{defi}[theorem]{Definition}
\newtheorem{ampli}[theorem]{Amplification}
\newcommand{\CD}{{\cal D}}

\newcommand{\CS}{{\cal S}}

\newcommand{\CN}{{\cal N}}
\newcommand{\CF}{{\cal F}}

\newcommand{\CG}{{\cal G}}
\newcommand{\CM}{{ M}}

\newcommand{\CL}{{\cal L}}

\makeatletter
\@addtoreset{equation}{section}

\makeatother
\newcommand{\RR}{{\mathbb R}}
\newcommand{\CC}{{\mathbb C}}

\newcommand{\ZZ}{{\mathbb Z}}

\newcommand{\NN}{{\mathbb N}}

\def\al{\alpha}

\def\la{\lambda}

\newcommand{\rref}[1]{(\ref{#1})} %puts parentheses around ref's

\newcommand{\del}{{\partial}}

\def\SyLe{$\CS_{\boldsymbol{\kappa}}$}

\def\mat2#1#2#3#4{{\left(\begin{array}{cc}#1 & #2\\ #3 & #4
      \end{array}\right)}}

\def\mats2#1#2#3#4{{\left(\begin{array}{cc}#1 & #2\vspace{2truemm} \\ #3 & #4 
\end{array}\right)}}

\def\ddd#1#2{\displaystyle{\frac{\partial #1}{\partial #2}}}

\newcommand{\Ha}[1]{H^{(#1)}}

\def\Nij{Nijenhuis}
\def\HJ{Hamilton--Jacobi\ }

\def\endpf{\begin{flushright}$\square$\end{flushright}}

\def\alg{{\mathfrak g}}

\def\var{manifold}

\def\bih{biham\-il\-tonian}
\def\varb{\bih\ \var}
\def\ham{Hamiltonian}

\def\dncoo{Darboux-Nijenhuis coordinates}

\begin{document}
\begin{flushright}
Ref. SISSA 90/2003/FM
\end{flushright}
%\vspace{0.8truecm}
\begin{center}
{\Large\bf 
Poisson Pencils, Integrability, \\ and Separation of Variables}\\
\vspace{1.truecm}
{\Large Gregorio Falqui}\\ \vspace{0.5truecm}
SISSA, via Beirut 2/4, \\
I--34014 Trieste (Italy)\\
falqui@fm.sissa.it
\end{center}
%{\bf Keywords:} {Hamilton-Jacobi Equations, Bihamiltonian Manifolds,
%Separation of Variables, Generalized Toda Lattices.}
\begin{abstract}\noindent
In this paper we will review a recently introduced method for solving 
the Hamilton-Jacobi equations by the method of Separation of Variables. 
This method is based on the notion of
pencil of Poisson brackets and on the bihamiltonian approach
to integrable systems. 
We will  discuss how  separability conditions  
can be intrinsically characterized  within such a 
geometrical set-up, the definition of the separation coordinates being
encompassed in the \bih\ structure itself.
We finally discuss these constructions studying in details a particular
example, based on a generalization of the classical Toda Lattice. 
\end{abstract}
\tableofcontents
\section{Introduction}
The study of the separability of the \HJ (HJ) equations associated with 
a given Hamiltonian function $H$ is a very  classical issue in
Mechanics, dating back to the foundational works of Jacobi, St\"ackel,
Levi-Civita and others. It has recently received a strong renewed 
interests thanks to its applications to the theory of integrable PDEs of KdV
type (namely, the theory of finite gap integration) and to the theory of
quantum integrable systems (see, e.g.,~\cite{DKN90,Sk95}).

As it is well known, the problem can be formulated as follows. 
Let $(M,\omega)$ be a $2n$
dimensional symplectic manifold, and let $(p_1,\ldots, p_n,q_1,\ldots, q_n)\equiv(\mathbf{p},\mathbf{q})$ be
canonical coordinates (in $U\subset M$), i.e., 
$\omega_{\vert_U}=\sum_{i=1}^n dp_i\wedge dq_i$.

The (stationary) HJ equation reads
\begin{equation}
  \label{eq:1.1}
  H(q_1,\ldots,q_n;\ddd{S}{p_1},\dots,\ddd{S}{p_n})=E\>.
\end{equation}
\begin{defi}
A complete integral
$S(\mathbf{q};\alpha_1,\ldots,\alpha_n)$ of HJ is a solution of \rref{eq:1.1},
depending on $n$ parameters $(\alpha_1,\ldots, \alpha_n)$ such that 
$Det{
\frac{\del^2 S}{\del{q_i}\del{\alpha_j}}}\neq 0$\\
$H$ is said to be {\em separable} in the coordinates
  $(\mathbf{p},\mathbf{q})$ if HJ admits an additively separated 
complete integral, that is, a complete integral of the form
\begin{equation}
  \label{eq:1.sepci}
  S(\mathbf{q};\alpha_1,\ldots,\alpha_n)=\sum_{i=1}^n 
S_i(q_i;\alpha_1,\ldots,\alpha_n).
\end{equation}
\end{defi}
In this paper we will focus on an equivalent `Definition--Theorem' of
separability, originally due to Jacobi and recently widely used by Sklyanin
and his collaborators. Let us consider an {\em integrable} Hamiltonian $H$,
that is, let us suppose that, along with $H=H_1$ we have further $n-1$ mutually
commuting integrals of the motion $H_2,\ldots H_n$, with $dH_1\wedge\ldots
\wedge dH_n\neq 0$.\\
\begin{defi}
An integrable system $(H_1,\ldots,H_n)$ is separable in the coordinates
$(\mathbf{p},\mathbf{q})$ if there exist $n$ non-trivial relations
\begin{equation}
  \label{eq:1.sjask}
  \Phi_i(q_i,p_i;H_1,\ldots,H_n)=0,\quad i=1,\ldots n\>,
\end{equation}
connecting {\em single pairs $(q_i,p_i)$} of canonical coordinates with the $n$
Hamiltonians $H_i$.
\end{defi}
We called this a Definition--Theorem due to the fact that it is a {\em
  constructive} approach to separability, since the knowledge of the
separation relations \rref{eq:1.sjask} allows one to reduce the 
problem of finding a
separated solution of HJ to quadratures. 
Indeed, one can solve the relations 
$\Phi_i(q_i,p_i;H_1,\ldots,H_n)=0,\quad i=1,\ldots n$ for the $p_i$ to get
$p_i=p_i(q_i;H_1,\ldots,H_n)$
and then define:
\begin{equation}
  \label{eq:1.sepsolu}
  S(\mathbf{q};\alpha_1,\ldots,\alpha_n)=\sum_{i=1}^n \int^{q_i}
  p_i(q'_i;H_1,\ldots,H_n)_{\big\vert_{H_i=\alpha_i}} dq'_i\>.
\end{equation}
This is by construction a separated solution of HJ; the fact that it is a
complete integral is equivalent to the fact that the integrals of the motion
depend non trivially on the momenta.
 
In intrinsic terms, one notices that, the equations 
$H_i=\alpha_i, \> i=1,\ldots, n$
define a foliation $\CF$ of $M$. The leaves of $\CF$ are
nothing but the (generalized) tori of the Arnol'd-Liouville theorem.
The foliation is {\em Lagrangian}, that is, the restriction of the 
two-form $\omega$ to $\CF$ vanishes. Hence the restriction to $\CF$
of the Liouville form
$\theta=\sum_{i=1}^n p_i d q_i$ to $\CF$ is (locally) exact. Indeed, the
function $S$ defined by \rref{eq:1.sepsolu} is  a (local) potential for
the restriction of $\theta$ to $\CF$. 
What is non intrinsic, and singles out the separation coordinates
$(\mathbf{p},\mathbf{q})$ is that the separation relations~\rref{eq:1.sjask},
which are another set of defining equations for the foliation $\CF$, 
have the very special property of containing a single pair of 
canonical coordinates
at a time. The problem to find such a system of coordinates and relations is
the core of the theory of SoV. In particular, a natural
question arises: \par
{\em
Is it possible to formulate intrinsic condition(s) on the
Hamiltonians $(H_1,\ldots,H_n)$ to {\em a priori} ensure 
separability in a set of canonical coordinates?}
\par
Actually, this is the main issue studied by both the `classical' 
Eisenhart-Benenti theory
of separability of natural systems defined 
on cotangent bundles to  Riemannian manifolds $(M, g)$, 
as well as the `modern' theory, due to the St. Petersburg and Montreal 
schools, of SoV for
systems admitting a Lax representation. We notice that both
such general approaches require the presence of {\em an additional structure}
to solve the problem. Indeed, the Eisenhart-Benenti theory requires the
existence of a conformal Killing tensor for the metric $g$, while the Lax
theory requires -- in addition to the knowledge of a Lax representation 
with spectral parameter for the Hamiltonian system under study -- the 
existence of an $r$--matrix structure for such a  Lax representation.

The method we will present/review in this paper, which has recently been
exposed in the literature (see,
e.g.,~\cite{BCRR96,MT97,Bl98,FMT98,Bl99,IMM00,FMPZ00,MT02,Pe02,FP03,BFP03}).
can be seen as a kind of
bridge between the classical and the modern points of view, putting an
emphasis on the geometrical structures of the Hamiltonian theory.
Its `additional'
structure is simply the requirement of the existence, on the symplectic
manifold $(M,\omega)$ of a {\em second} Hamiltonian structure, compatible with
the one defined by $\omega$. Namely, the \bih\ structure on $M$ will
allow us: 
\begin{enumerate}
\item To encompass the definition of a special set of coordinates, to be
  called {\em Darboux--Nijenhuis (DN)} coordinates, within a well defined
  geometrical object.
\item To formulate intrinsic (i.e., tensorial)
conditions for the separability of a Hamiltonian integrable system,
  in  the DN coordinates associated with the \bih\ structure.
\item To give recipes to characterize, find  and handle sets of
  DN coordinates. 
\end{enumerate}
The paper is organized as follows.
In Section \ref{sec:2} we will briefly introduce the notions of \bih\ geometry
relevant for the paper. In particular, we will discuss the notion of DN
coordinates, as well as methods to find them 
In Section \ref{sec:3} we will discuss the main theorems of the \bih\
set--up for SoV, namely, the tensorial conditions ensuring Separability of the
HJ equations  in DN coordinates.

Then we will discuss our
constructions in a specific example, whose separability, 
to the best of our knowledge, has
not yet been considered in the literature. It is a generalization of
the periodic Toda lattice with four sites.
In Section  \ref{sec:4} we will recall its definition, 
and show how  the ``\bih\ recipe'' for
SoV  can be applied to it. 
Although our constructions can be generalized to the generic $N$-site
system, for the sake of concreteness and brevity 
we choose to consider the four-site system only, 
and sometimes rely on direct computations to prove some of its properties.

In the last subsection we will apply our geometrical
scheme to 
study a specific reduction of this generalized Toda system, and
find integrals of the motion which are not encompassed in the Lax
representation. This result can be possibly  as first step towards
an alternative approach to the so--called chopping method 
of \cite{DLNT84}
for the full (non-periodic) Toda Lattice. 
\section{Some issues in the geometry of (bi)hamiltonian manifolds}\label{sec:2}
We start this Section recalling some well known facts in the theory of Poisson
manifolds.
\begin{defi}
A Poisson manifold $(M, \{\cdot,\cdot\})$ is a manifold endowed with a Poisson
bracket, that is a
bilinear antisymmetric composition laws defined on the space
$C^\infty(M)$ satisfying:
\begin{enumerate}
\item The Leibnitz rule: $\{fg,h\}=f\{g,h\}+g\{f,h\}$;
\item The Jacobi identity $\{f,\{g,h\}\}+\{g,\{h,f\}\}+\{h,\{f,g\}\}=0$.
\end{enumerate}
\end{defi}
We recall that a Poisson bracket (or Poisson structure) 
can be equivalently described with a
Poisson tensor, that is, with an application $P:T^*M\to TM$, smoothly varying
with $m\in M$ defined by:
\[
\{f,g\}=\langle df, P dg \rangle
\]
where $ \langle \cdot,\cdot \rangle$ denotes the canonical pairing between
$T^*M\text{ and } TM$. In a given
coordinate system $(x^1,\ldots,x^n)$ on $M$, the Poisson tensor $P$ associated
with the Poisson bracket $\{\cdot,\cdot\}$ is represented as
\[
P=\sum_{i,j} P^{ij}\ddd{}{x_i}\wedge\ddd{}{x_j},\quad\text{ with } 
P^{ij}=\{x^i,x^j\}.
\]
The Jacobi identity  is  translated into a quadratic differential 
condition on the
matrix $P^{ij}$ known as vanishing Schouten bracket (see, for further
details, e.g., \cite{MaMo84,Va96}), 
which in local coordinates reads
\begin{equation}\label{eq:sch0}
 \sum_{s=1}^n P^{is}\ddd{P^{jk}}{x_s}+  P^{js}\ddd{P^{ki}}{x_s}+
P^{ks}\ddd{P^{ij}}{x_s}=0, \> \forall\, i>j>k.
\end{equation}
We notice that
the existence of a Poisson structure does not give any restriction on the
dimension of $M$. As an example, the reader should consider the dual $\alg^*$ 
of a finite dimensional Lie algebra $\alg$, with Poisson brackets defined,
thanks to the natural identification  $\alg^{**}\simeq\alg $, by 
\[
\{f,g\}(X)=\langle X, [df(X),dg(X)]\rangle, \> X\in \alg^*, \> 
f,g\in C^\infty(\alg^*). 
\]     
The most familiar instances of Poisson manifolds
come from classical mechanics and are the cotangent bundles 
to  smooth manifolds. As it is well known (see, e.g.,~\cite{Ar89}),
they are equipped a canonical Poisson structure associated with 
a symplectic, that is, closed
(indeed exact) and non degenerate, two-form $\omega$. A Poisson manifold 
$(M, \{\cdot,\cdot\})$, is called {\em symplectic} if the Poisson tensor $P$
associated with $\{\cdot,\cdot\}$ is everywhere invertible.  
In such a case, the connection between the Poisson tensor $P$, the
Poisson brackets $\{f,g\}$ and the symplectic 
two--form $\omega$ can be described in the following equivalent ways:
\begin{description}
\item[a)] In local coordinates $(x_1,\ldots,x_n)\> $ one has
$\omega=\sum_{i,j} \omega_{ij} dx_i\wedge dx_j$,
where the matrix $\omega_{ij}$ is the inverse matrix to $P^{ij}$.
\item[b)] It holds the equality
\[
\{f,g\}=\omega(X_f,X_g)
\]
where, e.g., $X_f=P dF$ is the \ham\ vector field associated with $f$.
\item[c)]
For every vector field $Y$ and every function $f$, one has:
\[
L_Y(f)\equiv\langle df, Y \rangle=\omega(Y,X_f)
\]
\end{description}
The local structure of a Poisson manifold can be described as 
follows (see, e.g., \cite{We83, DKN90}).
Under suitable assumptions on the regularity of the Poisson brackets, one sees
that an open dense set of $M$ the Poisson tensor has rank
$r=2n$. There, $M$ is foliated in regular Poisson submanifolds, 
called generic symplectic leaves, that are
the (generic) common level sets of $k$ functions
$C_1,\ldots,C_k$, called Casimir functions of $P$. The distinguished property
of the Casimirs of $P$ is that their
Poisson bracket with any other function on
$M$ vanishes, or, equivalently, their differential lies in the kernel of $P$.
The dimension of $M$ is related with the integers $n$ and $k$ by $
\text{dim} M= k+ 2n$
As an example, we notice that, in the case of $M=\mathfrak{sl}(n)^*$, which we
can identify with $\mathfrak{sl}(n)$ itself 
by means, say,  of the trace form,
the Casimir functions are $C_i=\text{Tr}(X^{i+1}), i=1,\ldots, n-1$, and
the generic symplectic leaves consist of diagonalizable matrices with distinct
eigenvalues that can be identified with the direct sum of the lower and upper
nilpotent subalgebras. 

Let us now come to the definition of \varb.
\begin{defi}
A manifold $M$ is called a \varb\ if it is endowed with {\em two}
Poisson brackets $\{f,g\},\> \{f,g\}'$, such that,
for any $\la\in\RR$ (or $\la\in\CC$ if $M$ is complex), the linear
combination 
\begin{equation}\label{eq:2.poipen}
\{f,g\}'-\la \{f,g\}\equiv \langle df, (P-\la P') dg\rangle
\end{equation}
defines a Poisson bracket. 
This property is known as the {\em compatibility condition} between
the two brackets.
\end{defi}  
The expression  \rref{eq:2.poipen} will be referred to as 
{\em pencil} of Poisson brackets, and the sum 
$P_\la=P'-\la P$ pencil of Poisson tensors.
The most `popular' property of \varb s is contained in the
following
\begin{proposition}
Let $f$ and $f'$ two functions on a \varb\ $M$, which satisfy the
characteristic condition
$Pdf=P'df'.$
Then the Poisson brackets $\{f,f'\}$ and $\{f,f'\}'$ vanish.
\end{proposition}
{\bf Proof.}
It consists of a one--line computation. Let us consider, e. g., $\{f,f'\}$:
\[
\{f, f'\}=\langle df , P df'\rangle = - \langle df', P df \rangle = -\langle
df' P' df'\rangle =0.
\]
\endpf
\begin{defi}
A vector field $X$  that can be written as $X=Pdf=P'df'$
is called a {\em \bih\ vector field}. 
\end{defi}
%The fundamental
%process associated with a \bih\ structure is that of iteration. In fact the
%above Proposition admits the following  
\begin{corollary} \label{cor:1.1}
Let $f_k, k\in \ZZ$ be a sequence of (non-constant) functions satisfying
\begin{equation}\label{eq:1.bihrr}
Pdf_i=P'df_{i+1}. 
\end{equation}
Then
$ \{f_i, f_k\}=\{f_i,f_k\}'=0,\quad \forall\, i,k\in \ZZ.$
\end{corollary}
{\bf Proof.}
Using twice equation~\rref{eq:1.bihrr} and the antisymmetry of the Poisson
brackets we have, e.g.,
\[\begin{split}
&\{f_i,f_k\}=\langle df_i, Pdf_{k}\rangle = \langle df_i P' df_{k-1}\rangle=
-\langle df_{k-1},P' df_{i}\rangle=\\
&-\langle df_{k-1},P df_{i+1}\rangle=\langle df_{i+1},P
df_{k-1}\rangle=\{f_{i+1},f_{k-1}\}.
\end{split}
\]
Supposing $k>i$ and iterating this procedure $k-i$ times we get
$\{f_i,f_k\}=\{f_k,f_i\}$.
\endpf 

Following~\cite{GZ00}, we can state the following
\begin{ampli}\label{ampli:1.1}
Let $f_n, n=0,1,\ldots$ and $g_n,n=0,1,\ldots$ two sequences of (non-constant)
functions satisfying
\begin{equation}
  \label{eq:1.als}
  Pdf_i=P'df_{i+1}; Pdf_0=0,\quad Pdg_i=P'dg_{i+1}; Pdg_0=0.
\end{equation}
Then, along with $\{f_n, f_m\}=\{f_n, f_m\}'=\{g_n, g_m\}=\{g_n, g_m\}'=0$, it
holds
\[
\{f_n,g_m\}= \{f_n,g_m\}'=0,\quad \forall\, n,m\in \NN .
\]
\end{ampli}
The family of vector fields associated with a sequence of functions satisfying
the recursion relations~\rref{eq:1.bihrr} are customarily said to form a
Lenard-Magri sequence. Lenard-Magri sequences that start from the null vector
field, as in Amplification~\ref{ampli:1.1} are pictorially called {\em
  anchored} Lenard-Magri sequences. Notice that anchored
Lenard sequences can occur in \varb\ where at east one of the Poisson
brackets is non-symplectic (indeed, e.g., $df_0$ is a non-trivial 
element of the kernel of $P$). 
We can compactly express equations~\rref{eq:1.als}
relative, say,  to the sequence $f_i$ 
by considering the formal Laurent series
$f(\la)=\sum_{i=0}^\infty f_i/\la^i$ via the equation 
\begin{equation}\label{eq:cop}
(P'-\la P) df(\la)=0.
\end{equation}
In analogy with the definition of Casimir of a Poisson bracket,
Laurent series satisfying \rref{eq:cop} are called Casimirs
of the Poisson {\em pencil}. The reader should, however, 
bear in mind 
that while Casimir functions for a
single Poisson bracket are, in a sense, uninteresting functions,
Casimirs of a {\em pencil} of Poisson bracket compactly encode non-trivial
dynamics and constants of the motion.

If, as it often happens in the applications, 
inside the family $f_i$ we have an
element $f_n$ satisfying $P' df_n=0$, we can form a {\em polynomial} Casimir
of the pencil as %considering the finite sum
\begin{equation}\label{eq:2.polcas}
F(\la)=\la^nf_0+\la^{n-1}f_1+\cdots+f_n.
\end{equation}
Anchored Lenard sequences may give rise to families of integrable systems. 
Let us see how this happens in the case of a $2n+1$-dimensional
manifold endowed with a rank $2n$ pencil of Poisson tensors.
Let us suppose that we have found a polynomial Casimir of the
form~\rref{eq:2.polcas}, with the $n+1$ functions $f_0,\ldots, f_n$
independent. Let $\CS_c$ be a generic symplectic leaf of $P$ corresponding to
$f_0=c$. The vector fields $X_{f_i}, i=1,\ldots,n$ are tangent to $\CS_c$, are
Hamiltonian on $\CS_c$ (w.r.t. $P$) 
and the restriction of the functions $f_1,\ldots,f_n$
provide $n$ commuting integrals for each of them. In general it holds
\cite{GZ93,GZ00}:
\begin{proposition}\label{prop:gzint}
Let $(M,P,P')$ a \varb\ of dimension $d=2n+k$, and let $\text{dim}(Ker(P'
-\la P))=k$, for generic values of $\la$. Let us suppose that
$\Ha{1}(\la),\ldots, \Ha{k}(\la)$ are $k$ polynomial Casimirs of the pencil
$P_\la$ of the form
\[
\Ha{i}(\la)=\la^{n_i}\Ha{i}_0+\la^{n-1} \Ha{i}_1+\cdots  +\Ha{i}_{n_i}
\]
such that the collection of differentials $\{d\Ha{i}_j\}_{i=1,\ldots,k}^{j=0,\ldots
  n_i}$ defines a $k+n$ dimensional distribution in $T^*M$. Then the vector
fields defined by the anchored sequences associated with the $\Ha{i}$'s are
integrable (in the Arnol'd-Liouville sense) on the generic symplectic leaves
of $P$.  
\end{proposition}
\subsection{Geometry of regular \varb s and Darboux - Nijenhuis coordinates}\label{sub:2.1}     
An important class of \varb\ occurs when one element of the Poisson pencil
(which without loss of generality we will assume to be $P$) is everywhere
{\em invertible}, i.e., the 
Poisson bracket $\{\cdot,\cdot\}$ associated with $P$ is symplectic. 
The possibility of defining the inverse to one of the Poisson tensors lead us
to introduce a fundamental object in the \bih\ theory of SoV: 
the  \Nij\ (or Hereditary, or Recursion) operator 
\begin{equation}\label{eq:2.nij}   
N=P'\, P^{-1},
\end{equation}
(together with its 
dual $N^*= P^{-1}\, P'$). By definition,
$N$ (resp., $N^*$) is an endomorphism of the tangent bundle to $M$
(resp., of the cotangent bundle). 
As a remarkable consequence of the compatibility of $P$ and $P'$ the
\Nij\ torsion of $N$, defined by its action on a pair of vector fields $X,Y$ as
\begin{equation}\label{2.tn}
T(N) (X,Y) = [NX,NY] - N([NX,Y]+ [X,NY] - N[X,Y])
\end{equation} 
{\em identically vanishes} \cite{MaMo84}. So, from the classical
Fr\"olicher--\Nij\ theory, we know that its eigenspaces are {\em integrable}
distributions. Such distributions will be the building blocks of the
\bih\ set-up for SoV. 

To explain this, we have to make some remarks and a genericity assumption.
It can be shown that, owing to the antisymmetry of the Poisson tensors
defining $N$, pointwise the eigenspaces of $N$ are even dimensional. Througout
the paper, we will assume that for generic points $m\in M$, $N$ has the
maximal number $n=\frac12\text{dim}(M)$ of different eigenvalues
$\la_1,\ldots,\la_n$ , so that the
dimension of the eigenspace relative to any eigenvalue is $2$. Otherwise
stated, the characteristic polynomial of $N$ is the square of the 
degree $n=\frac12\text{dim}(M)$ minimal polynomial $\Delta_N(\la)$, whose
roots are pairwise distinct. We will call \varb s endowed with a Poisson pencil
with at least one of the elements of the Poisson pencil 
invertible, and such that the eigenvalues
of the associated \Nij\ tensor are maximally distinct,  {\em regular \varb s.} 
\begin{theorem} \label{th:2.6}On a regular 
%semisimple 
\varb\ there
  exist a class of coordinates $(y_i,x_i)$, 
to be called {\em Darboux--\Nij\ (DN)} coordinates,
  satisfying the two properties:
  \begin{description}
  \item[(Darboux)] They are canonical, that is, 
$\{x_i,y_j\}=\delta_{ij}$, $\{x_i,x_j\}=\{y_i,y_j\}$=0.
\item[(\Nij)] They diagonalize $N$, that is, $
N=\sum_{i} \la_i\big( \ddd{}{y_i}\otimes dy_i+\ddd{}{x_i}\otimes dx_i\big).$
  \end{description}
\end{theorem}
Th proof of this Theorem can be found in \cite{Ma90,GZ93}. 
Here we will sketch it and 
discuss its meaning. In words, the assertion states
that DN coordinates are defined by the spectral properties of $N$, as follows.
For all $m$ in the open set $U$ where the eigenvalues $\la_i$ of 
$N$ (which are the same as
the eigenvalues of $N^*$) satisfy $\la_i\neq\la_j, \> i\neq j$, the cotangent
space $T_m^*M$ admits the decomposition 
\begin{equation}\label{eq:3.deco}
T_m^*M=\oplus_{i=1}^n \CD_{m,\la_i}, \>\text{ dim } \CD_{m,\la_i}=2
\end{equation}
into eigenspaces of $N^*$. Thanks to the vanishing of the torsion of $N$, each
eigenspace $  \CD_{m,\la_i}$ is locally generated by differentials of pairs 
of
independent functions $(f_i,g_i)$.
This means that the pointwise decomposition~\rref{eq:3.deco} holds 
(in $U'\subset U$) as 
\[
T^*M_{\vert_U'}=\oplus_{i=1}^n \CD_{\la_i},
\]
where $\CD_{\la_i}$ is spanned by $df_i$ and $dg_i$,
with $N^* df_i=\la_i df_i$ and $N^* dg_i=\la_i dg_i.$%, i=1,\ldots, n.$

Functions whose differential belong to {\em
  different } summands $\CD_{\la_i}$ are in involution with respect 
to the Poisson brackets defined both by $P$ and $P'$.
Indeed, suppose that $f_1 \text{ and } f_2$ satisfy $N^*df_1=\lambda_1 df_1,
\> N^*df_2=\la_2 df_2$, $\la_1\neq \la_2$. The
relation $N^*=P^{-1}P'$ implies that
$
P'df_1=\la_1 P df_1,\> P'df_2=\la_2 P df_2.$
So,
\[ 
\{f_1,f_2\}'=\left\{\begin{array}{l} \langle df_1, P' df_2\rangle= \la_2 \langle
  df_1, P df_2\rangle =\la_2 \{f_1,f_2\}\\
         -\langle df_2, P' df_1 \rangle= -\la_1 \langle
  df_2, P df_1\rangle =\la_1 \{f_1,f_2\}\end{array}\right.
\]    
whence the assertion. It is equally straightforward to realize that
the only non vanishing Poisson brackets have the form
\[
\{f_i,g_i\}=F_i(f_i,g_i),\quad \{f_i,g_i\}'=F'_i(f_i,g_i), \quad i=1,\ldots, n.
\]
This means that, by quadratures,  from the 
$n$ pairs of  functions $\{f_i,g_i\}_{i=1,\ldots, n}$ 
we can construct a set of canonical coordinates satisfying the \Nij\ property
of Theorem~\ref{th:2.6}.
Thus the class of coordinates where to frame the \bih\ set-up for SoV admits a
clearcut and simple geometrical description. Admittedly, 
in the general case the computation of DN coordinates requires 
the integration of the two--dimensional distributions $\CD_{\la_i}$ 
associated with the eigenvalues $\la_i$ of $N^*$. Fortunately enough 
there are instances (that frequently occur in the
applications) in which DN coordinates can be found in an easier way.

\subsection{On \dncoo}\label{sub:3.2}
In this subsection we will briefly discuss conditions and `recipes' to 
algebraically find and/or characterize sets of \dncoo\ on regular \varb s. 
A very simplifying instance occurs whenever
the eigenvalues $\la_i$ of $N$ (that are, in general,
functions of the point $m\in M$) are {\em functionally independent}. 
It holds (see, e.g.,~\cite{MaMa96}): 
\begin{proposition}\label{prop:2.sdn} 
Let us define $I_i:=\frac{1}{2i}\text{Tr} N^i,\> i=1,\ldots, n$.
In the open set $U$ where $dI_1\wedge \cdots\wedge \cdots dI_n\neq 0$ 
the eigenvalues $\la_i\>, i=1,\ldots, n$ are functionally independent,
satisfy $N^*d\la_i=\la_i d\la_i$, and
so may be used to construct a set of \dncoo.
\end{proposition}
{\bf Proof.} We 
express the normalized traces $I_i$ of the \Nij\ tensor $N$ in terms of its
eigenvalues as $kI_k=\sum_{i=1}^n \la_i^k$. Hence
$dI_k=\sum_{i=1}^n \la_i^{k-1} d\la_i$,
that is, in matrix terms:
\begin{equation}\label{eq:2.vdm}
\left[\begin{array}{c}dI_1\\dI_2\\\vdots \\dI_n\end{array}\right]=
\left[\begin{array}{cccc} 
1&1&\cdots&1\\
\la_1&\la_2&\cdots&\la_n\\
\vdots&\vdots&\ddots&\vdots\\
\la_1^{n-1}&\la_2^{n-1}&\cdots&\la_n^{n-1}
\end{array}\right]\cdot
\left[\begin{array}{c}d\la_1\\d\la_2\\\vdots \\d\la_n\end{array}\right]
\end{equation}
So we have
\[
dI_1\wedge\cdots \wedge dI_n=\prod_{i\neq j} (\la_i-\la_j) 
d\la_1\wedge\cdots\wedge d\la_n,
\]
i.e., on the open set where the traces of the powers of the \Nij\ tensor are
functionally independent, we have that the eigenvalues $\la_i$ are different
and functionally independent. 

To proceed further we need to recall that \cite{MaMo84} that the 
normalized traces $I_i$ of the powers of \Nij\ operator
satisfy the recursion relation:
\begin{equation}\label{eq:2.durrI}
N^* dI_k=d I_{k+1}. 
\end{equation}
This can be proved as follows. At first one notices that
\rref{eq:2.durrI}  is equivalent to the relation
\[
L_{NX}(I_k)=L_X(I_{k+1}),\> \forall\>\text{vector field } X,
\]
as it can be easily seen evaluating the equality of one-forms
on a generic vector field $X$. Thanks to the Leibnitz property of the Lie
derivative and the ciclicity of the trace, we see that 
\begin{equation}\label{eq:2.trxn}
L_{NX}(I_k)=\text{Tr}(L_{NX}(N)\cdot N^{k-1})\quad\text{ and }
L_X(I_{k+1})=\text{Tr}(L_{X}(N)\cdot N^{k}).
\end{equation}
Since the vanishing of the \Nij\ torsion of $N$ implies that $
L_{NX}(N)=N\cdot L_{X}(N),\> \forall X $
the validity of \rref{eq:2.durrI} is proved..

We now express the relations~\rref{eq:2.durrI} 
in terms of the eigenvalues $\la_i$ 
\begin{equation}\label{eq:2.laex}
\left[\begin{array}{cccc} 
1&1&\cdots&1\\
\la_1&\la_2&\cdots&\la_n\\
\vdots&\vdots&\ddots&\vdots\\
\la_1^{n-1}&\la_2^{n-1}&\cdots&\la_n^{n-1}
\end{array}\right]\cdot
\left[\begin{array}{c}N^*d\la_1-\la_1d\la_1\\N^*d\la_2-\la_2 d\la_2\\
\vdots \\N^*d\la_n-\la_nd\la_n\end{array}\right]=
\left[\begin{array}{c}0\\0\\\vdots \\ 0\end{array}\right].
\end{equation}

Since the Vandermond matrix in the LHS of this equation is, 
by assumption, invertible, we conclude that, for
$i=1,\ldots, n$, it holds 
$N^*d\la_i=\la_i d\la_i$. 
\endpf
This proposition can be rephrazed saying that
``half of'' the DN coordinates
are algebraically provided by the \Nij\ tensor itself.
The remaining ``half'' 
$y_1,\ldots, y_n$ can always be found by quadratures. 
Actually, there is a
condition leading to the algebraic solution of this problem too. 
To elucidate this, the following two 
considerations are crucial.

First of all one remarks that, thanks to the fact that, for DN coordinates, 
if $i\neq j$ one has $\{\la_i,y_j\}=0$ one can replace the 
$n(n-1)/2$ equations
$\{\la_i,y_j\}=\delta_{ij}$ with the $n$ equations
\[
\{\la_1+\cdots+\la_n,y_j\}=1,\> j=1,\ldots, n,
\]
that do not involve the individual coordinates $\la_i$
but only their sum $\sum_{i=1}^n
\la_i=I_1$, and the Hamiltonian vector field 
\[
Y=-P dI_1=\sum_{i}\ddd{}{y_i}.
\]
The second argument goes as follows.  
Let us consider the distinguished functions $I_k$ introduced in 
proposition\ref{prop:2.sdn}, and trade them for the coefficients $p_i$ of the
minimal polynomial
\[
\Delta_N(\la)=\la^n-p_1\la^{n-1}-p_2\la^{n-2}-\cdots - p_n
\]
of $N$. The functions $p_k$ and $I_k$ are related by triangular Newton formulas
\begin{equation}\label{eq:3.new}
\begin{split}
&I_1=p_1;\quad I_2=p_2+\frac12 p_1^2;\quad I_3=p_3+p_2p_1+\frac13p_1^3;\\
&I_4=p_4+p_1p_3+p_1^2p_2+\frac12 p_2^2+\frac14 p_1^4;\quad 
I_5=p_5+\ldots.\end{split}
\end{equation}
As a consequence of the recursion relations~\rref{eq:2.durrI}, it can be easily
shown that the $p_i$'s satisfy the `Frobenius' recursion relations
\begin{equation}
  \label{eq:frrecrel}
  N^* dp_i=dp_{i+1}+p_i dp_1,\>\text{ with } p_{n+1}\equiv 0.
\end{equation}
We can compactly write these relations as a single relation for the polynomial
$\Delta_N(\la)$; indeed, a straightforward computation shows that they are
equivalent to
\begin{equation}
  \label{eq:deltarel}
  N^* d\Delta_N(\la)=\la d\Delta_N(\la)+\Delta_N(\la)d p_1\,.
\end{equation}
Actually relations of this kind are very important for our purposes. Indeed,
in \cite{FP03} we proved the following proposition:
\begin{proposition}\label{prop:nfg} 
Let $\Phi(\la)$ a smooth function defined on the manifold $M$, depending on
an additional parameter $\la$. Suppose that there exists a one-form
$\alpha_\Phi$ such that
\begin{equation}
  \label{eq:2.eeg}
  N^* d\Phi(\la)=\la d\Phi(\la)+\Delta_N(\la) \alpha_\Phi\>.
\end{equation}
Then, the $n$ functions $\Phi_i$ obtained evaluating the ``generating''
function $\Phi(\la)$ for $\la=\la_i, i=1,\ldots, n$ are \Nij\ functions,
that is, they satisfy $N^* d\Phi_i=\la_i d\Phi_i$.
\end{proposition}
\begin{defi} We will call a generating function $\Phi(\la)$ satisfying
  equation~\rref{eq:2.eeg} a {\em \Nij\ functions generator}. 
\end{defi}
The relevance of the notion of \Nij\ functions generator 
in the search for DN
coordinates stems from the fact that \Nij\ functions generators 
form an algebra $\CN(M)$, {\em which is closed} under the action
of the vector field $Y=-P dI_1$. 
%This menas that
%if $\Phi(\la)$ is a \Nij\ functions generator, so is $L_Y\Phi(\la)$. 
In this way, knowing a set of  \Nij\ functions generators, we can obtain
further elements of the algebra $\CN(M)$ by repeated applications of the
vector field $Y$. Clearly, in such an extended algebra, 
the characteristic equation 
\[
L_Y(\Psi(\la))=1+\Delta_N(\la)\alpha_\psi
\]  
may be easier to be solved, thus yielding
the missing \dncoo\ coordinates $y_i$ 
as $y_i=\Psi(\la_i).$
We will see an instance of this situation in the Example of Section~\ref{sec:4}.

For an analysis of \dncoo\ within the theory of multi-hamilronian structure on
loop algebras, see \cite{FMT98,DM02,HH02}
\section{The Separability Conditions}\label{sec:3}
As we have briefly recalled in Section \ref{sec:2}, on a \varb\  one is usually
lead to consider \bih\ vector fields, that is vector fields $X$ admitting
the twofold Hamiltonian representation $X=P df=P' dg$.
Let us now suppose that $(M, P, P')$ be a regular \varb\ of dimension $2n$,
and that we were able to construct, by means of the Lenard--Magri iteration
procedure a sequence of functions $H_1,H_2,\ldots$ satisfying 
$P'dH_i=P dH_{i+1}$. Let us also suppose that the first $n$ of them be
functionally independent. Then one easily shows that the all the further
Hamiltonians $H_{n+1},\ldots $
are functionally dependent from the first $n$. Indeed this
follows from the fact that a regular Poisson manifold of dimension $2n$  
cannot have more than $n$ mutually commuting independent functions.

This means that, if we consider the Hamiltonian $H_{n+1}$, there must be a
relation of the form
\begin{equation}\label{eq:2.relh}
\psi(H_1,\ldots, H_n;H_{n+1})=0,\quad\text{with } 
\psi_{H_{n+1}}\equiv\ddd{\psi}{H_{n+1}}\neq 0.
\end{equation}
relating it with the independent Hamiltonians $H_i, i=1,\ldots, n$.
 
Actually, the case of $H_i=I_i\equiv\frac1{2i} \text{Tr} N^i$ is an 
instance of this situation. In fact, since by the
Cayley--Hamilton theorem $N$ annihilates its minimal polynomial, we have
$ N^n-\sum_{i=1}^n p_i N^{n-1}=0$, yielding the relation
\[
I_{n+1}/{2(n+1)}-\sum_{i=1}^n 2(n-i+1) p_i I_{n-i+1}=0.
\]
%between the $n+1$ quantities $I_1,\ldots,I_{n+1}$.
Differentiating
equation~\rref{eq:2.relh} we see that, along with $P' dH_i=P dH_{i+1},
i=1,\ldots, n-1$ it holds:
\begin{equation}\label{eq:inv}  
P'dH_n=P dH_{n+1}=\frac{-1}{\psi_{H_{n+1}}} 
\big(\sum_{i=1}^n \ddd{\psi}{H_i} PdH_i\big),
\end{equation}
that is, the vector field $X_{n+1}=P dH_{n+1}=P'dH_n$ is a 
linear combination of the vector fields $X_1=P dH_1,\ldots X_n=P dH_n$. 

This innocent looking observation is the clue for the \bih\ theory of
SoV. Let $\{H_1, H_2,\ldots H_n\}$ any integrable system on $M$, that is 
that is, the $H_i$s are mutually commuting (w.r.t.$P$) independent functions. 
We can construct a $n$-dimensional distribution, namely
the distribution $\CD_{H}$ 
spanned by the $n$ mutually commuting vector fields $X_i=PdH_i$. This is
nothing but the very classical tangent distribution to the invariant tori of
the Liouville Arnol'd theory of integrable systems. 
Since $M$ comes equipped with a second  Poisson tensor $
P'$ we can as well consider the distribution $\CD'_{H}$
generated by the Hamiltonians $H_i$ under the action of $P'$, that is,
generated by the vector fields $X'_i=P'dH_i$.  It holds:
\begin{theorem}\label{th:2}
Let $\{H_1,\ldots,H_n'\}$ define, as explained above,  
an integrable system on  a regular \varb\  
$(M,P,P')$.
The Hamilton-Jacobi equations associated with  any of 
the Hamiltonians $H_i$ are
separable in the DN coordinates $x_,y_i$ defined by $N=P'P^{-1}$ 
if and only if the distribution $\CD'_{H}$ is contained in $\CD_{H}$, or,
equivalently, if the 
distribution $\CD_{H}$ is invariant along $N$. 
\end{theorem}
{\bf Proof.}
We will first prove the equivalence of the invariance of $\CD_H$ under $N$ and
the inclusion $\CD_H'\subset \CD_H$. To say that
$\CD'_{H}$ is contained in $\CD_{H}$ is tantamount to saying that there exists
a matrix $F_{ij}$, whose entries are, in general, functions defined on $M$
such that, for $i=1,\ldots, n$ it holds:
\begin{equation}\label{eq:sepcon}
X'_i\equiv P' dH_i=\sum_{j} F_{ij} P dH_j=\sum_{j} F_{ij} X_j.
\end{equation}
Writing $P'=NP$ we can translate these equalities into
$N X_i=\sum_{j} F_{ij} X_j$ for all $i=1,\ldots, n.$

The full proof of the fact that the invariance of $\CD_{H}$ 
insures separability in DN coordinates
can be found  in \cite{FP03}. It goes as follows.
%Here we will limit ourselves to sketch its main points.

At first we notice that the translation in terms of the 
codistribution $\CD^*_H$
generated by the differentials of the Hamiltonians $H_i$ of the invariance
condition for $\CD_H$ is the invariance condition $N^* \CD^*_H\subset
\CD^*_H.$ This can be easily seen applying to ~\rref{eq:sepcon} the operator 
$P^{-1}$, to get $N^* dH_i=\sum_{j} F_{ij} dH_j$.

Since all the Poisson brackets $\{H_i,H_j\}$ vanish and $M$ is a 
regular \varb,  
the matrix $F$ defined by \rref{eq:sepcon} can be show to have 
simple eigenvalues, that coincide with eigenvalues $\la_i$
of $N$. So there exists a matrix $S$ satisfying 
\[
SF=\Lambda S, \text{ where } \Lambda=\text{diag} (\la_1,\ldots, \la_n).
\]
If we define the $n$ one forms $\theta_i=\sum_{j} S_{ij}dH_j$, we get: 
\begin{equation}\label{eq:3.3}
N^* \theta_i=\sum_{j} S_{ij} N^* dH_j=\sum_{j,k} S_{ij}F_{j,k} dH_k=  
\sum_{j,k}\la_i\delta_{ij}S_{j,k}d H_k=\la_i \theta_i.
\end{equation}
This means that $\theta_i$ is an eigenvector of $N^*$ relative to $\la_i$;
hence there must be functions $F_i,G_i$ such that
\begin{equation}\label{eq:3.4}
\sum_j S_{ij} dH_j=F_i dx_i+G_idy_i
\end{equation}
whence the existence of a separation relation $\Phi_i(x_i,y_i; H_1,\ldots,
H_h)$ for all $i=1,\ldots, n$. The converse statement can be 
trivially proved.
\endpf
We would like to stress that%, as the attentive reader has already noticed,
the separability condition of Theorem \ref{th:2}
is a {\em tensorial one}. That is, given a regular \varb\  
$(M,P,P')$ this separability criterion can be checked in {\em any
  system of coordinates}, without the a--priori calculation of the DN
coordinates themselves. Notice, also, that the validity of the
statement does not (as it should be!) depend on the choice of mutually
commuting integrals $\{H_1,\ldots,H_n\}$. That is, if we consider a ``change
of coordinates in the space of the actions'', that is we trade the $H_i$'s for
another complete set of integrals of the motion $K_i=K_i(H_1,\ldots, H_n)$,
then the separability of the new Hamiltonians $K_i$ will hold if and only if
the separability of the original ones holds. Indeed, the dual distributions
generated by the $H_i$'s and the $K_i$'s coincide.

A second remark is important and deserves to be explicitly spelled out. 
Although we have started our discussion 
considering the case of a family of \bih\ vector fields, that is the case of
Lenard-Magri sequences, the hypotheses of Theorem \ref{th:2} concern only the
relations of the distributions generated respectively under the action of $P$
and $P'$ by the Hamiltonians $H_i$, {\em without any mention} of the fact that
the generators of the distribution be \bih\ vector fields. Thus, although it
might seem a somewhat odd statement, {\em the vector fields that are separable
  by means of the \bih\ approach are not necessarily \bih\ vector
  fields!}. It is also important to notice that it is not only a matter of 
choice of generators. Indeed, in \cite{TM96} it has been shown that
the only \bih\ vector fields on a regular \bih\ manifold turn out to be
associated with {\em separated functions of the eigenvalues of $N$}, that is 
functions of the form $H=\sum_{i=1}^n f_i(\la_i)$. This means that, in such a
case, the distribution $\CD_H$ coincides with that generated by the
distinguished functions $I_i$. However, this is by far a very special
example, that is, the range of applicability of the method is much wider than
that, as it as already been quite widely shown in the literature.

The separation condition of Theorem~\ref{th:2} is based on the analysis of the
behaviour of the characteristic distribution associated with an integrable
system under the \Nij\ tensor $N$. An equivalent criterion, based on the
analysis of the Poisson brackets associated with the tensor $P'$ can be
formulated as follows.
\begin{theorem}\label{th.3}  
Let $\{H=H_1, H_2,\ldots H_n\}$ an integrable system defined on regular 
\varb\  $(M,P,P')$.
The 
Hamiltonians $H_i$ are separable in the
DN coordinates defined by $N=P'P^{-1}$ 
if and only if, along with the commutation
relations $\{H_i,H_j\}=0$ there hold also
\begin{equation}
  \label{eq:3.5}
  \{H_i,H_j\}'\equiv\langle dH_i, P'dH_j\rangle =0,\> \text{ for } 
i,j=1,\ldots,n.
\end{equation}
\end{theorem}
{\bf Proof.} The key formula is the relation between $P,P'$ and $N^*$.
Indeed, suppose that $\CD_H^*$ be invariant along $N^*$. Then:
\[\begin{split}
 \{H_i,H_j\}'&=\langle dH_i, P'dH_j\rangle = \langle dH_i, NP dH_j\rangle \\
 & \langle N^*dH_i, P dH_j\rangle=\sum_k F_{ik} \langle 
dH_k, P dH_j\rangle=\sum_{k} 
F_{ik}\{H_k,H_i\}=0,\end{split}
\]
which, in view of Theorem~\ref{th:2} proves the statement in one direction.
Now, let us suppose that~\rref{eq:3.5} holds. Then, for every $ij=1,\ldots, n$
we have:
\[
0=\{H_i,H_j\}'=\langle dH_i, P'dH_j \rangle = \langle dH_i, NP dH_j \rangle=
\langle N^*dH_i, PdH_j \rangle,
\]
meaning that, for all $i=1,\ldots, n$, the one-form $N^*dH_i$ belongs to the
annihilator (w.r.t. $P$) of the distribution $\CD_H$. Since such an
annihilator coincides with $\CD^*_H$, this means that, for all $i=1,\ldots,
n$, $N^*dH_i\in \CD^*_H$.  
\endpf
This results leads to the following, somewhat daring, comparison. The
Liouville-Arnol'd theorem on finite dimensional integrable Hamiltonian systems
says that the geometrical structure underlying integrability of a Hamiltonian
vector field defined on a symplectic manifold $M,\omega$
is a {\em Lagrangian} foliation of $M$. We can rephrase the content of
Theorem~\ref{th.3} saying that the geometrical structure underlying the
separability of a system defined on a regular \varb $(M,P,P')$ is a {\em
  bilagrangian} foliation of $M$. 

We end our presentation of the \bih\ set-up for SoV with the following remark.
Theorem \ref{th:2} concerns only the existence of the separation relations. 
In principle, one could try to find these
relations in concrete examples by actually diagonalizing the matrix
$F$, and explicitly finding and integrating the
relations~\rref{eq:3.4}.
However, there is a very simple tensorial criterion 
which can be used to 
determine the {\em functional form} of the separation
relations  $\Phi_i(x_i,y_i; H_1,\ldots,
H_h)$, whose proof can be found in \cite{FP03}.
\begin{proposition}
Let $\{H_1,\ldots,H_n\}$ be an integrable system defined on a regular
\varb, which is separable in the \dncoo\
associated with $N=P^{-1}P'$. Consider the matrix
$F_{ij}$ fulfilling the relations~\rref{eq:sepcon}. Then the separation
relations are be {\em affine} in the Hamiltonians $H$, that is of the form
\begin{equation}\label{eq:stsep}
\Phi_i(x_i,y_i; H_1,\ldots,
H_h)
=\sum_{j} S_{ij}(x_i,y_i) H_j+U_i(x_i,y_i),
\end{equation}  
if and only if the matrix $F$ satisfies the relation $N^*dF_{ij}=\sum_{k} F_{ik} dF_{kj}.$
\end{proposition}
The matrix $S$ on ~\rref{eq:stsep} can be shown to be a suitably normalized
matrix of eigenvectors of the matrix $F$. 
Its characteristic property is that, as expressed in the
equation, 
the entries $S_{ij}$ of the $i$--th row depend only on the pair $(x_i,y_i)$ of
\dncoo. For this reason it can be called a
  St\"ackel matrix.
\section{Example: a generalized Toda Lattice}\label{sec:4}
In this final Section we will apply the general scheme outlined in the
previous Sections to a specific model, with the aim of showing how the recipes
discussed so far from a theoretical standpoint can be concretely applied. We
will study a  generalization of the four site Toda lattice,
to be termed {\em Toda$_3^4$ model}. This system is a member of a family 
introduced in \cite{Ku85} as  reductions of the discrete KP hierarchy.
It can be described as follows. We consider on
$\CM=\CC^{12}$, endowed with global coordinates $\{b_i,a_i,c_i\}_{i=1..4}$ the
Hamiltonian
\begin{equation}
  \label{eq:t:ham}
  H_{GT}=\frac12 (b_1^2+b_2^2+b_3+b_4^2)-(a_1+a_2+a_3+a_4),
\end{equation}
and the linear Poisson tensor given by the matrix
\begin{equation}
  \label{eq:4.p0}
  P=\left[\begin{array}{ccc}
\mathbf 0 &\mathbf {A_1} &\mathbf{C_1} \\
-\mathbf{A_1^T} & \mathbf{C_2}&\mathbf{0}   \\
-\mathbf{C_1^T}& \mathbf{0}&\mathbf{0} 
\end{array}\right]
\quad \text {with } 
\mathbf{A_1}= 
\left[ \begin {array}{cccc} -a_{{1}}&0&0&a_{{4}}\\\noalign{\medskip}
a_{{1}}&-a_{{2}}&0&0\\\noalign{\medskip}0&a_{{2}}&-a_{{3}}&0
\\\noalign{\medskip}0&0&a_{{3}}&-a_{{4}}\end {array} \right], 
\end{equation}
\begin{equation*}
 \mathbf{C_1}= \left[ \begin {array}{cccc} -c_{{1}}&0&c_{{3}}&0\\\noalign{\medskip}0
&-c_{{2}}&0&c_{{4}}\\\noalign{\medskip}c_{{1}}&0&-c_{{3}}&0
\\\noalign{\medskip}0&c_{{2}}&0&-c_{{4}}\end {array} \right], 
\quad \mathbf{C_2}= \left[ \begin {array}{cccc} 0&-c_{{1}}&0&c_{{4}}\\\noalign{\medskip}
c_{{1}}&0&-c_{{2}}&0\\\noalign{\medskip}0&c_{{2}}&0&-c_{{3}}
\\\noalign{\medskip}-c_{{4}}&0&c_{{3}}&0\end {array} \right].
\end{equation*}
where we denoted by $\mathbf{0}$ the $4\times 4$ matrix with
vanishing entries..\\
Using (here and in the sequel) the cyclic identifications
$a_{i+4}=a_i,b_{i+4}=b_i,c_{i+4}=c_i$,
the Hamiltonian vector field $X_{H_{GT}}=P d H_{GT}$ 
can be written as follows 
\begin{equation}
\label{eq:4xh}
\left[\begin{array}{c} \dot b_i\\\dot a_i\\\dot c_i\end{array}\right]=\left[
\begin{array}{l} a_{i-1}-a_i\\ a_i(b_{i+1}-b_i)+c_{i-1}-c_i\\
c_i(b_{i-2}-b_i\end{array}\right], \quad i=1,\ldots, 4,
\end{equation}
The expert reader will notice that $H_{GT}$ coincides with the
Hamiltonian of the periodic four-site Toda lattice, 
written in Flaschka coordinates
$b_i=p_i, a_i=\exp(q_i-q_{i+1})$. Indeed, on the hyperplane
$\CM_T\simeq \CC^8$ defined by $c_i=0, i=1,\ldots 4$ the vector field
$X_{H_GT}$ defines 
the periodic Toda flow.  
\begin{proposition}\label{prop:laxrep}
The Hamiltonian 
vector field $X_{H_{GT}}$ admits the Lax representation $\dot L(\mu) =
[L(\mu),\Phi]$, where
\begin{equation}
  \label{eq:4.lax}
  L(\mu)=  \left[ \begin {array}{cccc} b_{{1}}&-\mu&{\frac {c_{{3}}}{{\mu}^{2}}}
&{\frac {a_{{4}}}{\mu}}\\\noalign{\medskip}{\frac {a_{{1}}}{\mu}}&b_{{
2}}&-\mu&{\frac {c_{{4}}}{{\mu}^{2}}}\\\noalign{\medskip}{\frac {c_{{1
}}}{{\mu}^{2}}}&{\frac {a_{{2}}}{\mu}}&b_{{3}}&-\mu
\\\noalign{\medskip}-\mu&{\frac {c_{{2}}}{{\mu}^{2}}}&{\frac {a_{{3}}}
{\mu}}&b_{{4}}\end {array} \right],\qquad \Phi= \left[ \begin {array}{cccc} 0&0&{\frac {c_{{3}}}{{\mu}^{2}}}&{\frac {
a_{{4}}}{\mu}}\\\noalign{\medskip}{\frac {a_{{1}}}{\mu}}&0&0&{\frac {c
_{{4}}}{{\mu}^{2}}}\\\noalign{\medskip}{\frac {c_{{1}}}{{\mu}^{2}}}&{
\frac {a_{{2}}}{\mu}}&0&0\\\noalign{\medskip}0&{\frac {c_{{2}}}{{\mu}^
{2}}}&{\frac {a_{{3}}}{\mu}}&0\end {array} \right]. 
\end{equation}
\end{proposition}
The \bih\ aspects of this system have been discussed in \cite{Me01}. In
particular, it has been noticed that on $\CM$ there exists a second
Hamiltonian structure for the vector field $X_{H_{GT}}$. Namely one considers
the bivector $P'$ having the following form: 
\begin{equation}
  \label{eq:4.p1}
  P'=\left[\begin{array}{ccc} \mathbf{A_2}& \mathbf{B_1}& \mathbf{C_3}\\ 
-\mathbf{B_1^T}& \mathbf{A_2}
      &\mathbf{C_4}\\-\mathbf{C_3^T}&-\mathbf{C_4^T}&\mathbf{A_4}
\end{array}\right],\text { where } \mathbf{C_3}= \left[ \begin {array}{cccc} -b_{{1}}c_{{1}}&0&b_{{1}}c_{{3}}&0
\\\noalign{\medskip}0&-b_{{2}}c_{{2}}&0&b_{{2}}c_{{4}}
\\\noalign{\medskip}c_{{1}}b_{{3}}&0&-b_{{3}}c_{{3}}&0
\\\noalign{\medskip}0&c_{{2}}b_{{4}}&0&-b_{{4}}c_{{4}}\end {array}
 \right],
\end{equation}
\[
\mathbf{A_2}=
 \left[ \begin {array}{cccc} 0&a_{{1}}&0&-a_{{4}}\\\noalign{\medskip}-
a_{{1}}&0&a_{{2}}&0\\\noalign{\medskip}0&-a_{{2}}&0&a_{{3}}
\\\noalign{\medskip}a_{{4}}&0&-a_{{3}}&0\end {array} \right], 
\quad \mathbf{B_1}=
 \left[ \begin {array}{cccc} -b_{{1}}a_{{1}}&c_{{1}}&-c_{{3}}&b_{{1}}a
_{{4}}\\\noalign{\medskip}b_{{2}}a_{{1}}&-b_{{2}}a_{{2}}&c_{{2}}&-c_{{
4}}\\\noalign{\medskip}-c_{{1}}&b_{{3}}a_{{2}}&-b_{{3}}a_{{3}}&c_{{3}}
\\\noalign{\medskip}c_{{4}}&-c_{{2}}&b_{{4}}a_{{3}}&-b_{{4}}a_{{4}}
\end {array} \right],\]
\[
\mathbf{A_3}=  \left[ \begin {array}{cccc} 0&-b_{{2}}c_{{1}}-a_{{1}}a_{{2}}&0&b_{{1}
}c_{{4}}+a_{{1}}a_{{4}}\\\noalign{\medskip}b_{{2}}c_{{1}}+a_{{1}}a_{{2
}}&0&-b_{{3}}c_{{2}}-a_{{2}}a_{{3}}&0\\\noalign{\medskip}0&b_{{3}}c_{{
2}}+a_{{2}}a_{{3}}&0&-b_{{4}}c_{{3}}-a_{{3}}a_{{4}}
\\\noalign{\medskip}-b_{{1}}c_{{4}}-a_{{1}}a_{{4}}&0&b_{{4}}c_{{3}}+a_
{{3}}a_{{4}}&0\end {array} \right],\]
\[
\mathbf{C_4}= \left[ \begin {array}{cccc} -a_{{1}}c_{{1}}&-a_{{1}}c_{{2}}&a_{{1}}c_
{{3}}&a_{{1}}c_{{4}}\\\noalign{\medskip}c_{{1}}a_{{2}}&-a_{{2}}c_{{2}}
&-a_{{2}}c_{{3}}&a_{{2}}c_{{4}}\\\noalign{\medskip}c_{{1}}a_{{3}}&c_{{
2}}a_{{3}}&-a_{{3}}c_{{3}}&-a_{{3}}c_{{4}}\\\noalign{\medskip}-c_{{1}}
a_{{4}}&c_{{2}}a_{{4}}&c_{{3}}a_{{4}}&-a_{{4}}c_{{4}}\end {array}
 \right], \quad \mathbf{A_4}
= \left[ \begin {array}{cccc} 0&-c_{{1}}c_{{2}}&0&c_{{1}}c_{{4}}
\\\noalign{\medskip}c_{{1}}c_{{2}}&0&-c_{{2}}c_{{3}}&0
\\\noalign{\medskip}0&c_{{2}}c_{{3}}&0&-c_{{3}}c_{{4}}
\\\noalign{\medskip}-c_{{1}}c_{{4}}&0&c_{{3}}c_{{4}}&0\end {array}
 \right]. 
\]
It can be easily noticed that $X_{H_{GT}}=P' d(-\sum_{i=1}^4 b_i)$. More in
general, we have the following
\begin{proposition} The pencil $P'-\la P$ is a pencil of Poisson brackets.
The rank of the generic element of the pencil is eight. 
The characteristic polynomial $R(\la,\mu)=\text{Det}\big(\la\mathbf{1}-
L(\mu)\big)$,
written in terms of $\rho=\mu^4$ can be expanded as:
\begin{equation}
  \label{eq:4.cp}
  R(\la,\rho)=\la^4-\rho+H(\la)+({K(\la)-\la^2J_1})/{\rho}+{J_2}/{\rho^2}.
\end{equation}
The functions $J_1$ and $J_2$ are common Casimirs of $P$ and $P'$.
The polynomials $H(\la), K(\la)$ are polynomial Casimirs of the
pencil $P_\la=P'-\la P$. 
They have the form
\begin{equation}\label{eq:4.pcp}
H(\la)=\la^3H_0-\la^2 H_1+\la H_2-H_3, \> K(\la)=K_0\la +K_1.
\end{equation}
\end{proposition}
Explicitly, $J_1=c_1c_3+c_2c_4, J_2=c_1c_2c_3c_4$, while the coefficients 
of $H(\la)$ and $K(\la)$ are given by:
\[\begin{split}
&H_0=\sum_{i=1}^4 b_i,\> H_1=\sum_{i> j=1}^4 b_ib_j+\sum_{i=1}^4 a_i,\>
H_2=\sum_{i=1}^4 \big(c_i+b_i(a_{i+1}+a_{1+2})+b_ib_{i+1}b_{i+2}\big)\\
& H_3=\sum_{i=1}^4 b_ic_{i+1}+a_1a_3+a_2a_4 +\text{cubic and quartic terms};
\\
& K_0=\sum_{i=1}^4 b_i c_{i-1}c_{i+1}-c_i
a_{i-1}a_{i+2},\> K_1=\sum_{i=1}^4  a_ic_{1+1}c_{1+2} +
\text{quartic terms}\end{split}
\]
One can show that the eight functions $H_0,H_1,H_2,H_3,K_0,K_1, J_1,J_2$
are functionally independent and, thanks to the fact that they fill in Lenard
sequences, are mutually in involution. 
The kernel of $P$ is generated (at generic
points $m\in \CM$) by the differentials of the four functions
$H_0,K_0,J_1,J_2$. 
Hence, on the $8$ dimensional manifold $\CS_{\boldsymbol{\kappa}}$  
defined by the
equations $H_0=\kappa_1,K_0=\kappa_2,J_1=\kappa_3,J_2=\kappa_4$, that is, the
generic symplectic leaf of $P$, the vector field $X_{H_{GT}}$ is completely
integrable. To realize this we simply have to 
notice that $H_{GT}$ can be expressed as $\frac12 H_0^2-H_1$, and
apply the properties of  anchored Lenard Magri sequences collected in Proposition~\ref{prop:gzint}
 
\subsection{Separation of Variables}
We will now show how to apply the ideas and recipes if the \bih\ set-up 
for SoV to the Toda$_3^4$ model introduced above. The first problem to deal
with is that the Poisson tensor $P'$ does not restrict to \SyLe, but must be
projected. This can be done as follows (see \cite{DM02,FP02,FP03,MB03} 
for details and the
geometric background), by means of a kind of Dirac reduction process.

We consider the vector fields $Z_1=-\ddd{}{b_4}$ and $Z_2=\ddd{}{a_4}$,
notice that the matrix 
\[
G=\left(\begin{array}{cc} 
L_{Z_1}(H_0)&L_{Z_1}(K_1)\\ L_{Z_2}(H_0)&L_{Z_2}(K_1)\end{array}\right)= \left( \begin {array}{cc} 1&-c_{{1}}c_{{3}}\\\noalign{\medskip}0&-c_{
{1}}a_{{3}}-a_{{1}}c_{{2}}\end {array} \right) 
\]
is invertible, and form the bivector
\begin{equation}\label{eq:ult}
\Delta =\sum_{i,j=1}^2 [{G}^{-1}_{ij}] Z_i\wedge X_1^{j}, \text{ where }
X_1^1=P'd H_0, X_1^2=P'd K_0. 
\end{equation}
\begin{lemma}
The modified bivector 
%\begin{equation}\label{eq:4.Q}
$Q=P'-\Delta$
%\end{equation}
defines a Poisson bracket, compatible with $P$; $Q$ restricts to \SyLe. 
\end{lemma}
{\bf Proof.}
The proof of the fact that $Q_\la=Q-\la P$ is a Poisson pencil follows
(see, e.g., \cite{FP02}), from the equalities:
\begin{equation}\label{eq:4.trvf}
\begin{split}
&\qquad\qquad L_{Z_1} P=0,\quad L_{Z_1}P'=W^1_1\wedge
Z_1-c_3\ddd{}{a_3}\wedge Z_2 
\\ & L_{Z_2} P=(\ddd{}{b_1}-\ddd{}{a_1})\wedge Z_2,\quad
L_{Z_2}P'=(b_4\ddd{}{a_4}+\ddd{}{b_1})\wedge
Z_1+W^2_2\wedge
Z_2.\end{split}
\end{equation}
with $W^1_1= (a_3\ddd{}{a_3}-a_4\ddd{}{a_4}+c_2\ddd{}{c_2}-c_4\ddd{}{c_4})$
and 
\[
W^2_2=(b_4\ddd{}{b_4}-b_1\ddd{}{b_1}-a_3\ddd{}{a_3}-a_4\ddd{}{a_4}-c_1\ddd{}{c_1}+c_2\ddd{}{c_2}+c_3\ddd{}{c_3}-c_4\ddd{}{c_4}),
\]
as well as from the fact that
\begin{equation}\label{eq:4.qtos} 
Q dH_0=Q dK_0=dJ_1=Q dJ_2=0.
\end{equation}
To show that~\rref{eq:4.trvf} holds true 
is a matter of an explicit computation, while
\rref{eq:4.qtos} follows from the definition of $Q$. In fact, the last two
equations hold since $J_1$ and $J_2$ are Casimirs 
of $P'$ invariant under $Z_1$ and $Z_2$. 
For, e.g., $H_0$ one computes
\[\begin{split}
Q dH_0&=P'd H_0 -\Delta dH_0=
X^1_1-\sum_{i,j=1}^2 \big([{G}^{-1}_{i,j}]L_{Z_j}(H_0)\big)\cdot X_1^i\\ &\qquad=
X^1_1-\sum_{i,j=1}^2 \big([{G}^{-1}_{i,j}]{G}_{j1}\big)\cdot
X_1^i=X_1^1-\sum_{i}\delta_{i,1}\cdot X_1^i=0,
\end{split}
\]
where the second equality follows from the fact that all the functions
$H_i,K_\al,J_\al$ are in involution w.r.t. $P$.
\endpf
Thanks to the above lemma, the generic symplectic leaf \SyLe is endowed with
the structure of a regular \varb. It is easy to show that the non trivial
Hamiltonians $H_1,H_2,H_3, K_1$ (more precisely, the restriction to \SyLe of
these Hamiltonians) satisfy the hypothesis of Theorem \ref{th:2} w.r.t the
(restriction to \SyLe) of the pencil $Q-\la P$.
Indeed we have:
\[
Q dH_i=P'dH_i-\sum_{i,j=1}^2 [{G}^{-1}_{i,j}] Z_i\wedge X_1^{j}(dH_i)=
P dH_{i+1}-\sum_{i,j=1}^2 [{G}^{-1}_{i,j}] L_{Z_i}(H_i) X_1^{j}
\]
(where we understand $H_4=0$) and
\[
Q dK_2=P' dK_1-\sum_{i,j=1}^2 [{G}^{-1}_{i,j}] Z_i\wedge X_1^{j}(dK_2)=
\sum_{i,j=1}^2 [{G}^{-1}_{i,j}] L_{Z_i}(dK_1) X_1^{j}
\]

So we proved that, for generic values of the Casimirs
$\kappa_i, i=1,\ldots,4$, 
the system obtained by restriction of the Toda$^4_3$ flows
on \SyLe is separable in the DN coordinates associated with the restriction to
\SyLe of pencil $Q-\la P$. To finish our job we finally have to:
\\ \hspace{1cm} a)
explicitly define the DN coordinates; 
\\ \hspace{1cm} b)
find the separation relations.

To solve the first problem,  we will use the tools briefly
described in Section \ref{sub:3.2}.
We  rely on a result
of \cite{FP03}, as well as on explicit computations, 
to state the following
\begin{proposition}
Let $\CG(\lambda)$ be the matrix
\begin{equation}\label{eq:4g}
\CG=\left(\begin{array}{cc} L_{Z_1} H(\la)& L_{Z_2} H(\la)\\
 L_{Z_1} K(\la)& L_{Z_2} K(\la)\end{array}\right);
\end{equation}
The  roots of the degree $4$ polynomial
$\text{Det}(\CG)$ are the roots of the minimal polynomial
$\Delta(\la)=\la^4-\sum_{i=1}^4 p_i \la^{4-i}$ of the
\Nij\ tensor $N=P^{-1}Q$ associated with the regular Poisson pencil
$Q_\la$. The coefficients $p_i$ are well defined functions on the generic
symplectic leaf \SyLe,
and are functionally independent.
Furthermore, the ratios 
$\rho(\la)=-\CG_{2,2}/\CG_{1,2}, \sigma(\la)=-\CG_{2,1}/\CG_{1,1}$ 
are \Nij\ function generators.
\end{proposition}

Thus, one half of the \dncoo\ will be given by the roots of $\text{Det}(\CG)$.
To find the remaining half we consider vector field $Y={-P
dp_1}$ whose role has been discussed in Section \ref{sub:3.2}.
Since an explicit computations shows that
$L_{Y} \log(\rho(\la))=1$, hence we can state the following
%Hence, in view of the results of Section \ref{sub:3.2}
\begin{proposition}\label{prop:4.x}
A set of \dncoo\ for the restriction to the generic symplectic leaf \SyLe of
the Toda$^4_3$ flows are given by the four roots $\la_i$ of $
\text{Det}(\CG)(\la)$ and by the values $\mu_i$ of the function
$\log(\rho(\la))$ for $\la=\la_i$, where  
\[
\rho(\la)={\frac { \left( -c_{{1}}a_{{3}}-a_{{1}}c_{{2}} \right) \lambda+c_{{2}}
a_{{1}}b_{{3}}-a_{{1}}a_{{2}}a_{{3}}+c_{{1}}b_{{2}}a_{{3}}+c_{{1}}c_{{
2}}}{c_{{1}}c_{{3}}\lambda+a_{{1}}a_{{2}}c_{{3}}-c_{{1}}b_{{2}}c_{{3}}
}}
\]
\end{proposition}
To find the separation relations we reconsider the Lax matrix
~\rref{eq:4.lax}, and notice that the pairs, e.g., 
$\la_i,\rho(\la_i)$ are 
common solutions to the system
\begin{equation}\label{eq:4.linsys}
\begin{split}
&\rho \CG_{1,1}+\CG_{21}=0\\
&\rho\CG_{1,2}+\CG_{22}=0.
\end{split}
\end{equation}
Since the Lie derivative of the matrix $\CL(\la,\mu)=\la\mathbf{1}-L(\mu)$ 
along the vector fields $Z_i$
are given by:
\[
L_{Z_1}(\CL(\la,\mu))= 
\left[ \begin {array}{cccc} 0&0&0&0\\\noalign{\medskip}0&0&0&0
\\\noalign{\medskip}0&0&0&0\\\noalign{\medskip}0&0&0&1\end {array}
 \right], \>  
L_{Z_1}(\CL(\la,\mu))= 
 \left[ \begin {array}{cccc} 0&0&0&-{\mu}^{-1}\\\noalign{\medskip}0&0&0
&0\\\noalign{\medskip}0&0&0&0\\\noalign{\medskip}0&0&0&0\end {array}
 \right] 
\]
we see that the solutions of the system~\rref{eq:4.linsys} are nothing but the
solutions of the equations
\[
[\CL(\la,\mu)^\vee]_{44}=0,\> [\CL(\la,\mu)^\vee]_{41}=0,
\]
$\CL(\la,\mu)^\vee$ being the classical adjoint to $\CL(\la,\mu)=\la\mathbf{1}-L(\mu)$.
Using standard arguments of linear algebra, and the results collected in 
Proposition~\ref{prop:4.x}, we can state
\begin{proposition}
The separation relations connecting pairs of \dncoo\ $\la_i,\mu_i$, the
Hamiltonians $H_1,H_2,H_3, K_2$ and the Casimirs $H_0,K_1,K_2,J_0$ are, on the
generic symplectic leaf \SyLe, given by the evaluation of the characteristic
polynomial $\text{Det}(\CL(\la,\mu))$ in $\la=\la_i, 
\mu=\mu_i=\log(\rho(\la_i))$.
\end{proposition}
We notice that, {\em a posteriori,}  
the separation coordinates for the Toda$^4_3$ system
fall in the class described in, e.g.,\cite{Sk92,AHH93,DD94,Sk95,AHH97,KNS97}.  
Namely, the DN coordinates that separate
the Toda$^4_3$ system  are  Algebro-geometrical Darboux coordinates
associated with the spectral curve \rref{eq:4.cp}, and fulfill the so-called
Sklyanin's `magical recipe'.

As a final remark, in connection with the discussion on the relation between
the bihamiltonian property of an integrable vector field
and the separability of the associated HJ equations of Section \ref{sec:3}, 
we notice that the Hamiltonians $H_1,H_2,H_3,
K_2$ are functionally independent from the coefficients of the minimal
polynomial of the \Nij\ tensor obtained from $Q-\la P$. So, this is a further
instance of a system which is not \bih\ on a regular manifold, but 
turns out to be separable via the \bih\ method of SoV.
\subsection{A remarkable subsystem: the open Toda$^4_3$ system}\label{sec:5}
In this last subsection we will discuss a remarkable reduction
of the periodic Toda$^4_3$ system, leading to the corresponding generalization
of the open (or non-periodic) one. 
In the manifold $\CM\simeq \CC^{12}$ we consider the nine--dimensional
submanifold $\CM_0$ defined by the equations:
\begin{equation}
  \label{eq:5.ndef}
  a_4=c_3=c_4=0.
\end{equation}
One can easily verify that the restriction $X^0_{H_{GT}}$ to $\CM_0$ of the 
vector field $X_{H_{GT}}$ is tangent to $\CM_0$. Also, the tensor $P$ can be
restricted to $M_0$; indeed, the expression of its restriction 
$P_0$ w.r.t. the  natural coordinates
$\{b_1,\ldots,b_4, a_1,\ldots, a_3,c_1,c_2\}$ of $M_0$  is obtained from 
~\rref{eq:4.p0} simply by removing the 9th, 11th, 12th rows and columns.

One can easily check that $X^0_{H_{GT}}=P_0 d H_{GT,0}$ 
with
\begin{equation}\label{eq:h0}
H_{GT,0}=\frac12 (b_1^2+b_2^2+b_3+b_4^2)-(a_1+a_2+a_3),
\end{equation}  
and recognize that this function is the Hamiltonian of the open Toda lattice.
Also, a Lax pair for $X^0_{H_{GT}}$ is
\begin{equation}\label{eq:5.lap0}
  L_0=  \left[ \begin {array}{cccc} b_{{1}}&-\mu&0
& 0\\\noalign{\medskip}{\frac {a_{{1}}}{\mu}}&b_{{
2}}&-\mu& 0\\\noalign{\medskip}{\frac {c_{{1
}}}{{\mu}^{2}}}&{\frac {a_{{2}}}{\mu}}&b_{{3}}&-\mu
\\\noalign{\medskip}-\mu&{\frac {c_{{2}}}{{\mu}^{2}}}&{\frac {a_{{3}}}
{\mu}}&b_{{4}}\end {array} \right],
\qquad \Phi_0= \left[ \begin {array}{cccc} 0&0&
0&0
\\\noalign{\medskip}{\frac {a_{{1}}}{\mu}}&0&0&0\\
\noalign{\medskip}{\frac {c_{{1}}}{{\mu}^{2}}}&{
\frac {a_{{2}}}{\mu}}&0&0\\\noalign{\medskip}0&{\frac {c_{{2}}}{{\mu}^
{2}}}&{\frac {a_{{3}}}{\mu}}&0\end {array} \right]. 
\end{equation}

It should be clear from the form of the Lax pair that the vector field
$X^0_{H_{GT}}$ on $\CM_0$ is an extension of the standard open 
Toda lattice towards
the so-called full open Toda lattice, 
which is a system describing a flow on
the lower Borel subgroup of $sl(N)$. The integrability of the full open
Toda lattice was established in \cite{DLNT84} (see, also, 
\cite{EFS93}). The idea was to complement the integrals of the motion
coming from the Lax representation with additional integrals obtained
by means of the `chopping method', whithin the 
group--theoretical point of view.

The need to supply the standard results of the Lax theory 
with further methods should be
clear from the following considerations. The only Casimir
function of $P_0$ is $h_0=\sum_{i=1}^4 b_i$. 
Hence, its symplectic leaves
$\CS_\xi\subset(\CM_0)$ are the eight dimensional 
manifolds defined by
$h_0=\xi$, and $X^0_{H_{GT}}$ can be seen as 
a Hamiltonian system with {\em four}
degrees of freedom.

The characteristic polynomial of the matrix $L_0$ is
\begin{equation}
  \label{eq:5.cp0}
  \text{Det}(\la\mathbf{1}-L_0(\mu))=-\mu^4 +{\lambda}^{4} -h_0\la^3+h_1\la^2-h_2\la+h_3,
\end{equation}
that is, %apart from the Casimir $h_0$, 
it  provides us with only {\em three} non trivial Hamiltonians
\begin{equation}\label{eq:5.hh}\begin{split}
&h_1=\sum_{i>j=1}^4 b_ib_j+\sum_{i=1}^3 a_i,\quad 
h_2=\sum_{i>j>k=1}^4 {b_ib_jb_k}+\sum_{i=1}^3 a_i(b_{i+2}+b_{i+3})+c_1+c_2\\
& h_3=b_{{1}}a_{{2}}b_{{4}}+a_{{1}}b_{{3}}b_{{4}}+b_{{1}}b_{{2}}a_{{3}
}+b_{{1}}c_{{2}}+a_{{1}}a_{{3}}+c_{{1}}b_{{4}}+b_{{1}}b_{{2}}b_{{3}}b_
{{4}}.\end{split}
\end{equation}

We will now show how the the tools we previously introduced can be used
to geometrically prove the complete integrability of such a system and, 
moreover, yield the existence of an additional 
integral of the motion .

The main property is that, along with $P$, the tensor $Q$ resticts to $M_0$.
This can be proven as follows: one checks by direct inspection that this is
true for $P'$; the assertion follows from the fact that the vector field
$X^2_1$ (defined in ~\rref{eq:ult}) vanishes on $M_0$, 
while $Z_1$ and $X^1_1$, which coincides with $X_{GT}$,
are tangent to $M_0$ at the points of $M_0$. 

Furthermore, we add
two observations. The first one concerns the
restriction $\CG_0$ to $\CM_0$ of the matrix $\CG$. 
It has the form
\begin{equation}
  \label{eq:5.cg0}%\begin{split}
  \CG_0= \left[ \begin {array}{cc} \CG^0_{11}&{\lambda}^{2}- \left( b_{{2}}+b_{{3}}
 \right) \lambda+b_{{2}}b_{{3}}+a_{{2}}\\\noalign{\medskip}0& -\left( 
c_{{1}}a_{{3}}+a_{{1}}c_{{2}} \right) \lambda-a_{{1}}a_{{2}}a_{{3}}+c_
{{1}}c_{{2}}+c_{{1}}b_{{2}}a_{{3}}+a_{{1}}c_{{2}}b_{{3}}\end {array}
 \right]
\end{equation}
with $\CG^0_{11}=\la^3-\pi_1\la^2-\pi_2\la-\pi_3$ a degree three polynomial.
Hence its determinant (that is, the minimal polynomial of the \Nij\ tensor
$N_0$ induced by the pair $Q_0-\la P_0$ on $\CS_\xi$) factors 
as $\CG^0_{11}\CG^0_{22}$.

The second observation consists in the fact that the three surviving
Hamiltonians $h_1,h_2,h_3$ given by ~\rref{eq:5.hh} satisfy the conditions:
\begin{equation}\label{eq:5.recrelh0}
Q_0 dh_i=\sum_{j=1}^3 F_{ij}^0 P_0 dh_i, \text{ with } 
F_{ij}^0=\left[\begin{array}{ccc} \pi_1&1&0\\ \pi_2&0&1\\
    \pi_3&0&0\end{array}\right].
\end{equation}
We notice that the functions $\pi_i, i=1,\ldots, 3$ and the root 
\begin{equation}\label{eq:5.la4}
\la_4={\frac {-a_{{1}}a_{{2}}a_{{3}}+c_{{1}}c_{{2}}+a_{{1}}c_{{2}}b_{{3}}+c_
{{1}}b_{{2}}a_{{3}}}{c_{{1}}a_{{3}}+a_{{1}}c_{{2}}}}
\end{equation}
of $\CG^0_{2,2}$ are still functionally independent and hence (generically)
different on $\CS_\xi$. 
\begin{lemma} \label{lem:5.sig}
Let $\sigma$ any function satisfying
$Q_0 d\sigma=\la_4 P_0 d\sigma.$
Under
the above hypotheses, the brackets $
\{\sigma, h_i\}_{P_0}\text {  and } \{\sigma, h_i\}_{Q_0}$
vanish. 
\end{lemma}
{\bf Proof.} Evaluating both sides of $Q_0 d\sigma=\la_4 P_0 d\sigma$
on the differentials $(dh_1,dh_2, dh_3)$, and switching 
the action of the Poisson tensors on the $dh_i$'s, we get 
\[
\langle d\sigma, Q_0 dh_i\rangle= \la_4 \langle d\sigma, P_0 dh_i\rangle,
i=1,\ldots, 3
\]
Inserting~\rref{eq:5.recrelh0} we get the equation $
\sum_{j=1}^3 \big(F_{ij}^0-\la_4\delta_{ij}\big) \langle d\sigma, P_0
dh_j\rangle=0.$
Since $\la_4$ is {\em not} an eigenvalue of 
$F_{ij}^0$, the Lemma is proved.
\endpf
So a  fourth integral of the motion, that commutes with the Hamiltonian
$H^0_{GT}$  for the open Toda$^4_3$ lattice
is given indeed by the distinguished root $\la_4$ of equation~\rref{eq:5.la4};
this constructively proves the integrability of the system.

Finally, we notice that this method proves the existence of 
a {\em fifth} integral of the motion. Indeed, we know that, along with
$\la_4$, there must exist another independent 
function $\mu_4$, satisfying the hypotheses
of Lemma~\ref{lem:5.sig} and functionally independent of $\la_4$ and
of the  $h_i$'s. In such a comparatively low
dimensional case, such a function can be explicitly found to be 
\[
\mu_4={\frac {c_{{2}} \left( a_{{1}}b_{{2}}c_{{1}}-a_{{2}}{a_{{1}}}^{2}-{c_{
{1}}}^{2}-c_{{1}}b_{{3}}a_{{1}} \right) }{ c_{{1}}\left( c_{{1}}a_{{3}}+a_{{1
}}c_{{2}} \right) {(\lambda_{{4}}}^{3}-\pi_{{1}}{\lambda_{{4}}}^{2}-\pi_{{2}}\lambda_{{4}
}-\pi_{{3}})
}}.
\]
\section*{Acknowledgments}
I wish to thank Franco Magri for stimulating my interest in the theory of  
\varb s, and, in particular for sharing with me his insights about the 
problem of SoV. Most of the results herewith presented 
have been obtained in a long--standing
joint-work with him and Marco Pedroni, whose constant collaboration has
been invaluable for me. 
I am also grateful to 
Boris Dubrovin and Giorgio Tondo for a number useful discussions. 
Some computations  were performed with Maple $V^\circledR$.
This work has been partially supported by the
Italian M.I.U.R. under the research project {\em Geometry of
Integrable Systems.}

\end{document}